# A detailed first-principles study of the structural, elastic, thermomechanical and optoelectronic properties of binary rare-earth tritelluride NdTe$_3$

Tanbin Chowdhury[1], B. Rahman Rano[1*], Ishtiaque M. Syed[1], S. H. Naqib[2*]

[1]Department of Physics, University of Dhaka, Dhaka-1000, Bangladesh

[2]Department of Physics, University of Rajshahi, Rajshahi-6205, Bangladesh

*Corresponding authors' e-mail:  B. Rahman Rano - <rano167@du.ac.bd> and S. H. Naqib - <salehnaqib@yahoo.com>

**Abstract**

Rare-earth tritellurides ($R$Te$_3$) are popular for their charge density wave (CDW) phase, magnetotransport properties and pressure induced superconducting state among other features. In this literature, Density functional theory has been exploited to study various properties of NdTe$_3$. The calculated elastic and thermomechanical parameters, which were hitherto untouched for any $R$Te$_3$, uncover soft, ductile, highly machinable and damage tolerant characteristics, as well as highly anisotropic mechanical behavior of this layered compound. Its thermomechanical properties make it a prospective thermal barrier coating material. Band structure, density of states, Fermi surfaces and various optical functions of the material have been reported. The band structure demonstrates highly directional metallic nature. The highly dispersive bands indicate very low effective charge carrier mass for the in-plane directions. The Fermi surfaces display symmetric pockets, including signs of nesting, bilayer splitting among others, corroborating previous works. The optical spectra expose high reflectivity across the visible region, while absorption is high in the ultraviolet region. Two plasma frequencies are noticed in the optical loss function. The optical conductivity, reflectivity and absorption reaffirm its metallic properties. The electronic band structure manifests evidence of CDW phase in the ground state.

## I. Introduction

Recently rare-earth tritellurides ($R$Te$_3$) have gained tremendous interest due to their charge density wave (CDW) phase [1–5], magnetotransport properties [6,7], pressure induced superconducting state [8,9], and competition between CDW and magnetically ordered phases [10]. Large number of studies experimentally detected the CDW phase in $R$Te$_3$ using angle-resolved photoemission spectroscopy (ARPES) [11–13] and quantum oscillations [14,16] among other methods. Some of these investigations, including others, contain theoretical sections on $R$Te$_3$ compounds using first-principles and tight binding methods [7,12,17]. They show promising physical aspects to design electronic states by magnetic field and ultra-short light pulse [16,18,19]. Moreover, they turned out to be an ideal system for studying the Higgs (amplitude) mode of the CDW by quantum interference methods [20]. These are related to novel electronic ground states of $R$Te$_3$ and their complex interplays.

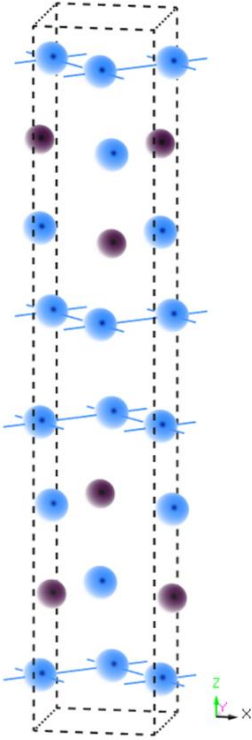

Figure 1: Crystal structure of NdTe$_3$. The brown balls represent Nd atoms and the blue balls represent Te atoms.

All $R$Te$_3$ crystallize in *Bmmb* space group with an orthorhombic column-like structure [21,22], each conventional unit cell consisting of four molecular units as showed in Fig. 1. The "layered" structure basically comprises of alternate stacking of a dual layer of almost square planes made of Te sheets, and a combined $R$Te layer, which is one $R$ and one Te layer intertwined into one cubic NaCl-type structure [21,23,24]. A weak interlayer bonding holds these stacked $R$Te and double Te layers together. The Te sheets are believed to be separately responsible for electric conduction, while the $R$Te layers dominate magnetic properties [10,24]. This makes the electrons roam without restraint in two dimensions, but they are restricted within the square planes, just like graphene [16,25]. If $R$ in $R$Te$_3$, goes from lighter to heavier metals across the lanthanide series in the periodic table, the lattice parameters decrease, i.e., replacing $R$ constitutes as chemical pressure [26]. This chemical pressure can modify the CDW properties (CDW gap, CDW transition temperature $T_{CDW}$, CDW wave vector) without having to introduce defect [27,28]. Again, an amazing feature of these materials is that applying external pressure on one of the lighter $R$Te$_3$ is equivalent to chemical pressure when it comes to the aforementioned properties [28], which implies modification in the Fermi surface (FS) under applied pressure [10]. While lighter members ($R$) of $R$Te$_3$, including NdTe$_3$, show one CDW phase, heavier ones have more than one [27,29].

$R$Te$_3$ are characterized by partially gapped region in FS, which originates from quasi one-dimensional $p_x$ and $p_y$ orbitals in the square Te lattice [30]. The FS consists of a diamond shaped pocket in the middle of the Brillouin zone (BZ), and four outer pockets near the zone boundary [12]. It has been proposed that the gap opens up due to displacement of the diamond pocket arisen from the Te layers by a nesting wave vector, while others argue that electron-phonon coupling is most likely in charge [30–32]. Both FS and quantum oscillations (QO) enrich us crucially when it comes to CDW transition. An FS study on NdTe$_3$ unearthed excellent features using both ARPES and tight binding calculations [30]. The authors successfully solved the otherwise unsolved problem of assigning QO frequencies to FS elements, and hence explained high mobility of charge carriers. Previously, a QO study on NdTe$_3$ probed low temperature magnetotransport properties using both de Haas–Van Alphen (dHvA) and Shubnikov–de Haas (SdH) effects [16]. The authors showed that mobility of electrons in NdTe$_3$ increases by 58% below the magnetic-ordering temperature ($T_N$), which is due to mitigation of spin-disorder scattering disrupting antiferromagnetic order [33-35]. Another anomaly in temperature-dependent resistivity around 20 K is noted when an external magnetic field was applied. The temperature dependent SdH oscillation curve also has anomalies below T$_N$ and around 20 K i.e., it deviates from the conventional Lifshitz-Kosevich fit [16].

Other studies on $R$Te$_3$ include electrical resistivity, and magnetic susceptibility along two crystallographic axes, with varying temperature [23]. The authors reported anisotropy in magnetic susceptibility, and anomaly in both susceptibility and resistivity for NdTe$_3$. Another paper contributed with temperature-dependent specific heat among other properties [24]. While the former study clearly pointed out susceptibility deviates from Curie-Weiss law at low temperature, both studies report two closely separated 2$^{nd}$ order phase transition for NdTe$_3$ (more for another $R$Te$_3$ compound). In addition, an experimental optical spectroscopy study of NdTe$_3$ including comparative analysis of all $R$Te$_3$ revealed evidence of CDW transition [36]. Electronic band structure, density of states, phonon dispersion, Raman spectra and Raman active modes of NdTe$_3$ have also been reported in the literature [31,37]. For other $R$Te$_3$ compounds, it was found that the optoelectronic features can be tuned using chemical and external pressure [38] and possibly by intercalating with Pd atoms [39] or introducing Te vacancies [40]. Some of the $R$Te$_3$ compounds are not environmentally very stable; stability decreases with heavier $R$ as 4$f$ electrons increase, which have a direct impact on degeneration [10,41]. But luckily in degenerated $R$Te$_3$, CDW properties remain intact.

It is surprising to note that, although many of the exotic electronic, magnetic and CDW features of NdTe$_3$ have been explored both theoretically and experimentally in detail, the more conventional theoretical study on the elastic, thermal, bonding and optical properties have not been done exhaustively yet. All these hitherto unexamined bulk properties are extremely important to check the viability of potential applications of this compound. In this particular work, using first-principles method, elastic and thermomechanical properties of NdTe$_3$ have been studied, which are not reported for any $R$Te$_3$, to the best of our knowledge. It was found that the compound under interest is a soft, ductile, highly machinable and damage tolerant system, showing highly anisotropic mechanical behavior. It is a potential thermal barrier coating agent, and it possesses optical characteristics to be used as solar reflector and efficient ultraviolet radiation absorber. We reaffirm highly directional metallic character and low effective charge carrier mass in the in-plane directions of NdTe$_3$. We have revisited the FS and found various pockets, signs of nesting, bilayer splitting and other features. We presented various optical functions at different energies and detected two plasma frequencies. We have identified evidence of CDW phase as well.

The rest of this article is organized as follows: Part II contains an overview of methodology. In Part III, along with a discussion on optimized structure, aforementioned bulk properties are reported and compared with those of the other literatures. We have included comparative analysis with other $R$Te$_3$ compounds whenever possible. Then part IV is for the concluding remarks.

## II. Computational Scheme

Density functional theory (DFT) has been used throughout the paper, which means solving for the ground state electron energy using the Kohn-Sham equation [42]. Computations have been carried out with local density approximation (LDA) exchange-correlation functional [43] using the quantum mechanical CAmbridge Serial Total Energy Package (CASTEP) [44]. As far as the electron-ion interactions are concerned, norm-conserving pseudopotentials and Koelling-Harmon relativistic treatment were used [45,46] to achieve high level of accuracy. For self-consistent calculations, density mixing electronic minimizer was chosen. Broyden-Fletcher-Goldfarb-Shanno (BFGS) algorithm was implemented to optimize the crystal geometry [47]. The electronic configurations used for pseudo atomic calculations are, Nd: [$4f4\ 5s2\ 5p6\ 6s2$] and Te: [$5s2\ 5p4$]. In order that final enthalpy saturates during optimization, a special *k*-point sample of size 12x12x2 in the Brillouin zone based on the Monkhorst-Pack scheme [48] and a plane wave basis set cut-off energy of 550 eV have been employed. The convergence thresholds set up while optimizing the geometry are: $10^{-5}$ eV atom$^{-1}$ for the energy, 0.03 eV Å$^{-1}$ for the maximum force, 0.05 GPa for maximum stress and $10^{-3}$ Å for maximum atomic displacement. A spin polarized calculation has been performed, since Nd has parallel 4*f* electron spins. To optimize the spin states, a sufficiently large number of empty bands were added. Spin orbit coupling (SOC) was not included because it was reported that for *R*Te$_3$ systems, it generates an insignificant effect on bulk physical characteristics [31,39,40,49]. Furthermore, our calculations without SOC agrees well with the prior experimental results.

To calculate the single-crystal elastic constants $C_{ij}$, and elastic compliances $S_{ij}$, the stress-strain technique from the CASTEP package has been utilized. From these elastic constants, all the other elastic moduli and thermomechanical parameters have been determined using well established methodology [50-52]. When it comes to optical functions, at first, the complex dielectric function, $\varepsilon(\omega) = \varepsilon_1(\omega) + i\varepsilon_2(\omega)$, was realized. Taking advantage of the Kramers-Kronig transformations, the real part $\varepsilon_1(\omega)$, of the dielectric function was derived from the imaginary part $\varepsilon_2(\omega)$. From the matrix elements of the electronic transitions between occupied and unoccupied electronic states, $\varepsilon_2(\omega)$ was calculated using the formula programmed into CASTEP which is given by:

$$\varepsilon_2(\omega) = \frac{2e^2\pi}{\Omega\varepsilon_0} \sum_{k,v,c} |\langle \Psi_k^c|\hat{u}.\vec{r}|\Psi_k^v\rangle|^2 \delta(E_k^c - E_k^v - E)$$

where $\Omega$, $\omega$, $e$ and $\hat{u}$ are the unit cell volume, the incident photon's angular frequency, electron's charge and is a unit vector denoting the incident electric field polarization, respectively, while $\Psi_k^c$ and $\Psi_k^v$ are the wave functions of conduction and valence electrons, respectively, with wave-vector *k*. This equation describes vertical optical transitions which is the dominant channel in the ground state. The purpose of the delta function is to ensure the conservation of energy and momentum during the optical transition process. The other five optical functions are derived from the complex dielectric function using well-known formulae [53-55].

## III. Results and Analysis
### A. Structure

The space group of NdTe$_3$ is found to be *Bmmb* (no. 63), or equivalently *Bm2$_1$b*, or *B2mb*. All the crystallogaphically independent atoms reside in position 4c, x = 0 and y = 0.25 in the plane, while the z-axis coordinates are 0.8306 (Nd), 0.0705 (Te), 0.4294 (Te) and 0.7047 (Te) [21]. The optimized lattice constants are tabulated below (Table 1). We have also included previously obtained experimental lattice parameters in this Table. Reasonable agreement has been found. It should be noted that the experimental results were obtained at finite temperature (close to room temperature), while our theoretical values are appropriate at 0 K. It must also be kept in mind that some authors orient the longer axis along c-direction, while others along b; we have adopted the former and changed the notations where needed.

Table-1: The lattice parameters a, b, c (Å) and volume V (Å$^3$) of NdTe$_3$.

|  | a | b | c | V | Volume deviation (%) |
|---|---|---|---|---|---|
| Experimental [21] | 4.350 | 4.350 | 25.800 | 488.201 |  |
| Experimental [22] | 4.347 | 4.363 | 25.851 | 490.276 |  |
| This work | 4.315 | 4.313 | 25.565 | 475.755 | 2.55% – 2.57% |

There are two types of layers in the structure: Te layers and NdTe combined layers. The arrangement of 8 layers along c-axis is as follows, Te, NdTe, NdTe, Te, Te, NdTe, NdTe, Te. There are 4 Nd and 12 Te atoms in a unit cell. This (for any $R$Te$_3$) arrangement is in fact two slightly distorted NdTe$_2$ (any $R$Te$_2$) unit cells stacked over one another, with an added Te layer between the cells, with every other cell translated by a/2, which opens up interesting opportunity for comparison [21]. Some authors have worked on polytellurides in general as well [39]. The compound NdTe$_3$, thus, forms an incommensurate structure [22,26].

### B. Elastic Properties

Elastic constants are the generalized coefficients of Hooke's law, which in general form a 6x6 matrix, also referred to by the name the stiffness matrix or the elastic matrix. But due to symmetry considerations, elastic matrix of an orthorhombic crystal has only nine independent components: $C_{11}$, $C_{22}$, $C_{33}$, $C_{44}$, $C_{55}$, $C_{66}$, $C_{12}$, $C_{13}$, and $C_{23}$. For NdTe$_3$, the independent elastic constants $C_{ij}$, and elastic compliance constants $S_{ij}$, determined using the LDA are given in Table 2.

Table-2: The single crystal elastic constants $C_{ij}$ (GPa), and elastic compliance constants $S_{ij}$ (GPa$^{-1}$) of NdTe$_3$.

| ij | $C_{ij}$ | $S_{ij}$ |
|---|---|---|
| 11 | 92.128 | 0.0237387 |
| 22 | 100.419 | 0.0218200 |
| 33 | 37.706 | 0.0299205 |
| 44 | 13.740 | 0.0727783 |
| 55 | 13.166 | 0.0759573 |
| 66 | 57.902 | 0.0172707 |
| 12 | 70.374 | -0.0159857 |
| 13 | 18.405 | -0.0033697 |
| 23 | 19.383 | -0.0034137 |

Among the nine stiffness constants, $C_{11}$, $C_{22}$ and $C_{33}$ quantify the crystal's capacity to withstand applied mechanical stress along the crystallographic a-, b-, and c-axis, respectively. For NdTe$_3$, $C_{33}$ is way smaller than $C_{11}$ and $C_{22}$, pointing out that it is way more compressible in the c-direction than the other two. This is expected since bonding in the in-plane directions would be stronger due to its layered nature. The ability of a crystal to resist shear is measured by the constants $C_{44}$, $C_{55}$ and $C_{66}$. Shearing strains dictate mechanical failure mode of crystalline solid when the stress is tangential to crystal plane. $C_{44}$ also gives an idea about the indentation hardness of a solid. For NdTe$_3$, small values of $C_{44}$ and $C_{55}$ indicate its limited ability to counter shear in (100) and (010) planes, respectively. Large value of $C_{66}$ compared to $C_{44}$ and $C_{55}$ implies that [100](001) shear is harder than the other two shears. The off-diagonal elements, $C_{12}$, $C_{13}$, and $C_{23}$ measure the defiance against volume conserving orthogonal distortions. The large gap between $C_{11}$ and $C_{33}$ is manifested in the lowest value of $C_{13}$. This shows the amount of stress produced along the crystallographic a-direction due to the uniaxial strain along crystallographic c-direction. The same pattern follows for $C_{12}$ and $C_{23}$. The mechanical stability of a solid can be tested from the elastic constants using the Born-Huang criteria [56]. For orthorhombic crystals, the appropriate Born-Huang criteria in the ground state are given by [57]:

$$C_{11} > 0; C_{11}C_{22} > C_{12}^2$$

$$C_{11}C_{22}C_{33} + 2C_{12}C_{13}C_{23} - C_{11}C_{23}^2 - C_{22}C_{13}^2 - C_{33}C_{12}^2 > 0$$

$$C_{44} > 0; C_{55} > 0; C_{66} > 0$$

All the criteria are fulfilled by the elastic constants of NdTe$_3$; therefore, this material is mechanically stable.

Table-3: The isotropic bulk modulus $B$ (GPa), and shear modulus $G$ (GPa), for polycrystalline NdTe$_3$ using the single crystal elastic constants according to Voigt, Reuss and Hill's methods, and the Pugh ratio $B/G$, Young's modulus $E$ (GPa), Poisson's ratio $\upsilon$, and the machinability index $\mu_M$, according to Hill's approximation.

| $B_R$ | $B_V$ | $B_H$ | $G_R$ | $G_V$ | $G_H$ | $B/G$ | $E$ | $\upsilon$ | $\mu_M$ |
|---|---|---|---|---|---|---|---|---|---|
| 33.399 | 49.619 | 41.509 | 16.835 | 25.101 | 20.968 | 1.979 | 53.838 | 0.284 | 3.021 |

Using the single crystal elastic constants and compliances, polycrystalline elastic moduli can be found [58]. The obtained polycrystalline bulk modulus $B$, shear modulus $G$, Pugh's ratio $B/G$, Young's modulus $E$, Poisson's ratio $\upsilon$, and machinability index $\mu_M$, using various approximations are presented in Table 3. The assumption in Voigt approximation is continuous strain and discontinuous stress distribution, and consequently there is a lack of balance in actual stresses among the grains [59]. While, conversely, in Reuss approximation, continuous stress and discontinuous strain distribution are assumed, therefore, strained grains do not fit smoothly [60]. This is why the former approximation delivers the maximum value of polycrystalline elastic moduli and the latter gives the minimum. On the other hand, Hill's technique uses the arithmetic mean of the two values, which is closer to the actual value of elastic moduli [61]. The bulk modulus and the shear modulus determine the resistance to volume change due to applied isotropic pressure and plastic deformation because of shear, respectively. These values are comparatively small for NdTe$_3$, proving that it is a soft and damage tolerant material.

The Pugh's ratio, Young's modulus, Poisson's ratio and the machinability index can be found from the bulk and shear moduli [58]. Since, $B > G$ for NdTe$_3$, its mechanical failure mode is governed by shear. Large value of the Pugh's ratio (> 1.75) means ductility and small value (< 1.75) means brittle material. Therefore, NdTe$_3$ is a ductile compound according to our computations. By definition, Young's modulus is the ratio of the tensile stress to the longitudinal strain, which determines the stiffness of a material. Very small value of $E$ for NdTe$_3$ implies that it is a less stiff solid. The failure mode of solids can be unraveled from Poisson's ratio as well. If $v$ is greater (less) than 0.26, the material is ductile (brittle) [62]. NdTe$_3$ is a ductile material consistent with the prediction from Pugh's ratio. Since, $0.25 < v < 0.50$ for NdTe$_3$, the interatomic forces are expected to be dominated by central forces [63]. For pure covalent bonding, $v$ is around 0.10, while a value of around 0.33 means metallic bonding, which implies combination of both types of bonds in NdTe$_3$. The machinability index, given by the ratio of bulk modulus to $C_{44}$, holds significance in the materials design industry [64]. The large value of $\mu_M$ for NdTe$_3$ means very good level of machinability.

Table-4: The calculated values of hardness $H_i$, i=1-8 (GPa) of NdTe$_3$

| $H_1$ | $H_2$ | $H_3$ | $H_4$ | $H_5$ | $H_6$ | $H_7$ | $H_8$ |
|---|---|---|---|---|---|---|---|
| 3.997 | 3.268 | 3.093 | 3.419 | 0.810 | 2.329 | 3.022 | 2.335 |

The resistance to permanent deformation, namely the hardness, is a very important feature for industry. The hardness $H_i$, has been calculated from the bulk, shear and Young's moduli using various semi-empirical formulae [65–68], for NdTe$_3$, and are listed in Table 4. These values are close to each other and are quite low.

### C. Elastic Anisotropy

Elastic anisotropy is about directional dependence of mechanical properties. It contains information about several physical properties important for material design, such as creation of micro-cracks, motion of cracks, formation of plastic deformations in crystals, etc. From the elastic moduli and constants, various anisotropy factors for NdTe$_3$ have been calculated [58,69,70] and tabulated in Table 5. Shear anisotropy factors are measures of anisotropy in the bonding between atoms with respect to various crystal planes. $A_1$ is the shear anisotropy factor for the {100} shear planes between the <011> and <010> directions, $A_2$ is for the {010} shear planes between the <101> and <001> directions, and $A_3$ is the factor for the {001} shear planes between the <110> and <010> directions. $A_1 = A_2 = A_3$ would mean isotropic crystal with respect to shape deformation. Any value greater or less than one would point out the degree of anisotropy, which is very high for NdTe$_3$, as expected due to its highly layered structure. The log-Euclidean index is zero for perfectly isotropic crystals. Materials with higher (lower) $A_L$ values exhibit strongly layered (non-layered) structure [70]. Although $A_L$ can be as high as 10.26, $A_L < 1$ for 90% of the crystals. A comparatively high value verifies layered structure of NdTe$_3$. A zero value of the universal anisotropy index also indicates absolute isotropy. The deviation from zero measures the degree of anisotropy.

Table-5: Shear anisotropy factors $A_1$, $A_2$ and $A_3$, universal log-Euclidean index $A_L$, the universal anisotropy index $A_U$, equivalent Zener anisotropy measure $A_{eq}$, and anisotropy in shear $A_G$ and anisotropy in compressibility $A_B$, for NdTe$_3$.

| $A_1$ | $A_2$ | $A_3$ | $A^L$ | $A^U$ | $A^{eq}$ | $A_G$ | $A_B$ |
|---|---|---|---|---|---|---|---|
| 0.591 | 0.530 | 4.471 | 0.405 | 0.683 | 2.091 | 0.197 | 0.195 |

For an isotropic crystal, $A^{eq} = 1$. A very high value indicates highly anisotropic behavior. Range of anisotropy in shear and compressibility is from 0 to 1. A zero value means perfect isotropy and unity means maximum possible anisotropy. The computed values of $A_G$ and $A_B$ are high for NdTe$_3$.

Table-6: Uniaxial bulk moduli $B_a$, $B_b$ and $B_c$ (GPa), anisotropy indices for bulk modulus $A_{Ba}$ and $A_{Bc}$, and for linear compressibility $\beta_a$ and $\beta_c$ (TPa$^{-1}$), along different crystallographic axes, and ratio of the compressibility $\beta_c/\beta_a$, of NdTe$_3$.

| $B_a$ | $B_b$ | $B_c$ | $A_{Ba}$ | $A_{Bc}$ | $\beta_a$ | $\beta_c$ | $\beta_a/\beta_c$ |
|---|---|---|---|---|---|---|---|
| 228.139 | 413.124 | 43.221 | 0.552 | 0.105 | 3.542 | 23.063 | 0.154 |

The uniaxial bulk modulus along a-, b- and c-axis, anisotropies of the bulk modulus along a- and c-axis with respect to the b-axis, and compressibility along a- and c-axis have been calculated [58,71] and listed in Table 6. Strong anisotropy is evident in the uniaxial bulk modulus. These values are way larger than isotropic bulk modulus. Linear compressibility along crystallographic c-axis is significantly higher than that along a-axis, due to the layered structure of NdTe$_3$. Small value of $\beta_a/\beta_c$ is yet another measure of its marked anisotropy.

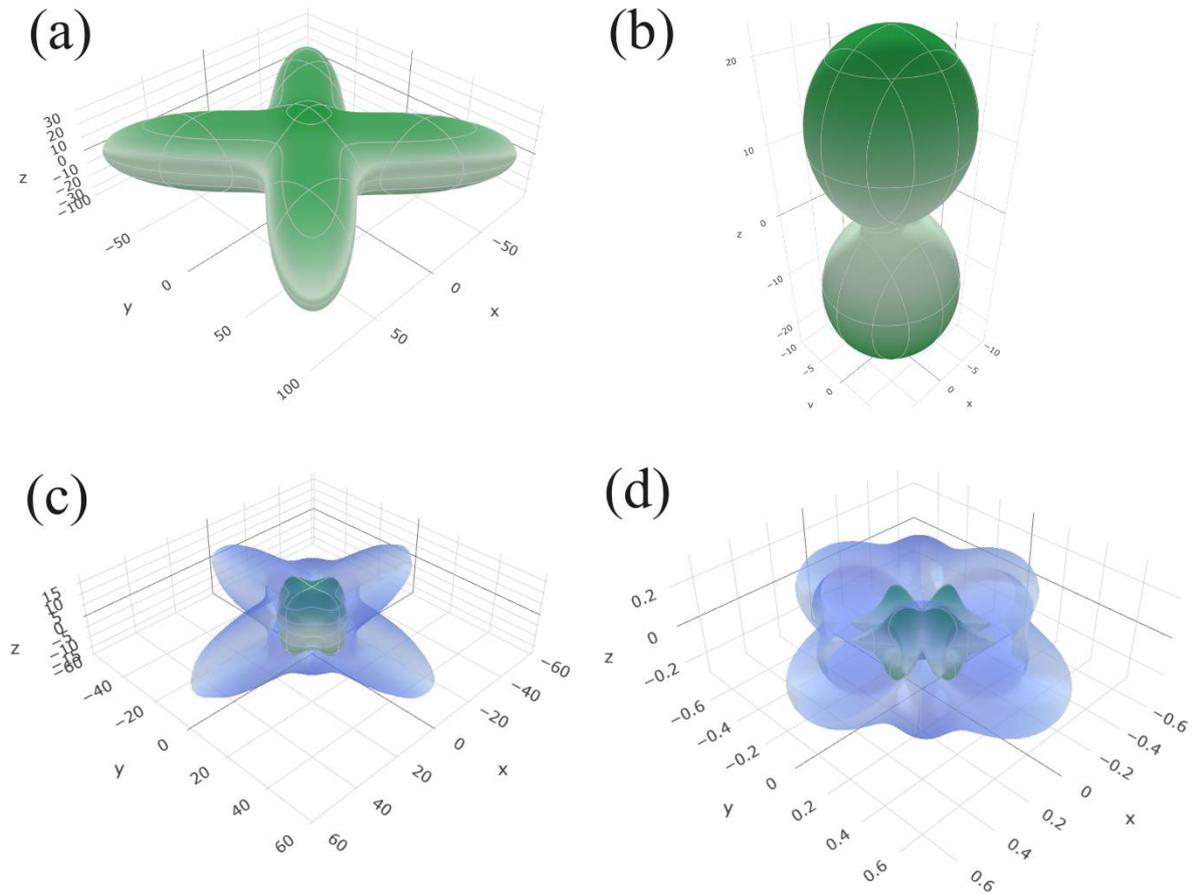

Figure 2: Three dimensional directional dependences of (a) Young's modulus (b) compressibility (c) shear modulus and (d) Poisson's ratio for NdTe$_3$. Green and blue surfaces represent lower and higher limits of the parameters, respectively.

Three dimensional (3D) plots of Young's modulus, compressibility, shear modulus and Poisson's ratio using the ELATE code [72], have been plotted for NdTe$_3$ in Fig. 2. Green and blue surfaces represent lower and higher limits of the parameters, respectively. For an isotropic crystal the profiles would be spherical. On the other hand, the more distorted the shapes, the more anisotropic they are. Almost all the profiles show extreme anisotropy. The code also shows projections of these surfaces along three crystal planes, with circular shapes implying isotopy. As expected from high degree of anisotropy, none of the projections are anywhere near circular.

### D. Thermomechanical Properties

The Debye temperature $\theta_D$, melting temperature $T_m$, Grüneisen parameter $\gamma$, and minimum phonon thermal conductivity $k_{min}$, have been calculated from elastic constants and moduli employing well-established theoretical formalisms [58,73–75] and are tabulated in Table 7. All these parameters are significant to explore the potential of a solid in various thermomechanical applications. The Debye temperature is a very important physical parameter, related to thermal conductivity, lattice vibration, superconducting transition temperature, interatomic bonding strength, melting temperature, coefficient of thermal expansion, phonon specific heat etc. Compounds with smaller Debye temperature have weaker interatomic bond strength, higher average atomic mass, lower melting temperature, low hardness and lower acoustic wave velocity. According to Debye model, $\theta_D$ corresponds to the highest allowed phonon frequency of a single normal mode vibration in a crystal. For NdTe$_3$, the Debye temperature calculated is small, revealing its softness and low thermal conductivity at lower temperature.

Table-7: The Debye temperature $\theta_D$ (K), melting temperature $T_m$ (K), Grüneisen parameter $\gamma$, and minimum phonon thermal conductivity $k_{min}$ (Wm$^{-1}$K$^{-1}$) for NdTe$_3$.

| $\theta_D$ | $T_m$ | $\gamma$ | $k_{min}$ |
|---|---|---|---|
| 180.785 | 686.943 | 1.677 | 0.271 |

Like the Debye temperature, $T_m$ of NdTe$_3$ is not high, indicating low overall bond strength, consistent with other parameters discussed previously. The dimensionless quantity Grüneisen parameter is an assessment of lattice anharmonicity, large value of $\gamma$ implying greater anharmonicity. The compound NdTe$_3$ possesses medium level of anharmonicity. It points out temperature-dependence of the phonon frequencies and their damping, as well as how phonon frequency is linked to variation of volume due to anharmonicity in the lattice potentials. It is also useful for determining thermal expansion effects, thermal conductivity and phase transitions related to volume change. At high temperatures above $\theta_D$, the thermal conductivity approaches to a minimum value ($k_{min}$). This occurs when the phonon mean-free path becomes lower than the average interatomic separation. Generally, compounds with lower $\theta_D$ such as NdTe$_3$ have lower $k_{min}$. The estimated value of the minimum thermal conductivity is quite low. This, together with very low Debye temperature and high damage tolerance of NdTe$_3$, implies that this can serve as a thermal barrier coating (TBC) material [68].

## E. Electronic Properties
### a) Band Structure

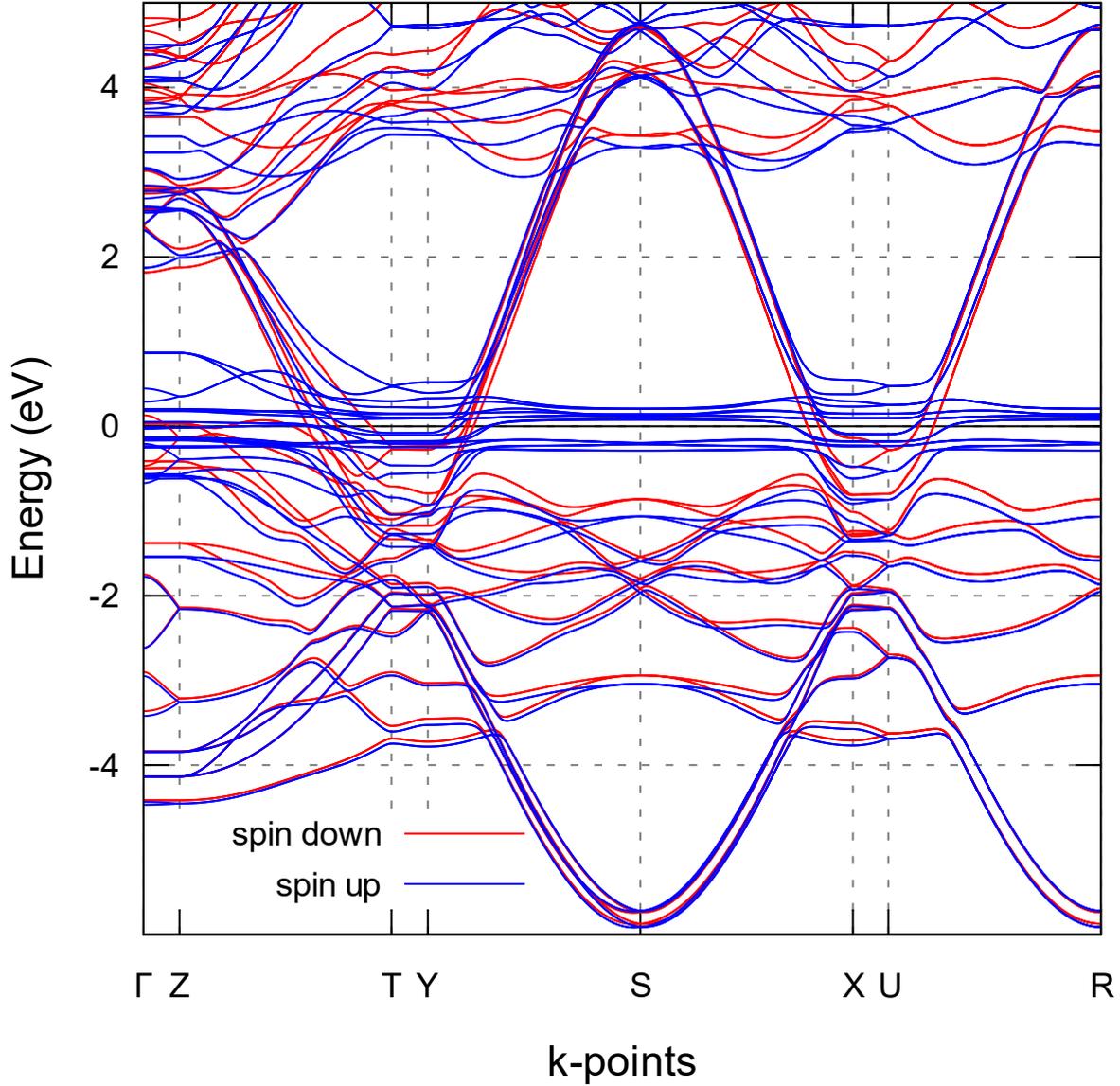

Figure 3: The spin-resolved electronic band structure of NdTe$_3$ along the high symmetry directions in the Brillouin zone.

The electronic band structure (BS) of NdTe$_3$ along high symmetry directions in the Brillouin zone (BZ) is illustrated in Fig. 3. The Fermi level $E_F$, is set at 0 eV. Several highly dispersive bands cross $E_F$, indicating a metallic behavior. While bands along c-axis (Γ-Z, T-Y and X-U) are mostly non-dispersive, many of the bands in the ab-plane (Z-T, Y-S-X and U-R) are exceptionally dispersive, which means that effective mass of the charge carriers in ab-plane is much smaller compared to those running along the c-axis. This clearly reveals highly directional metallic characters. This is anticipated since electrical conductivity is dominated by Te atom containing planes [10,21,24]. In isostructural SmTe$_3$, the in-plane conductivity is more than 3000 times higher than the out-of-plane value [76]. We predict similar anisotropic feature for NdTe$_3$ as well. Therefore, the highly dispersive bands that cross $E_F$ must be the Te in-pane $p$ orbitals, and less dispersive bands crossing $E_F$ might be due to the localized Nd 4$f$ electronic orbitals. These are the electrons responsible for strong electronic correlations and magnetic order in NdTe$_3$. This assumption as well as the shape of our BS is consistent with previously calculated result for NdTe$_3$ [31]. This is further confirmed by the electronic dispersion curves of LaTe$_3$ and LuTe$_3$, where there is no partially filled *f* orbital and therefore no less dispersive bands close to $E_F$; but the similar dispersive bands are still present [17,31]. There is notable effect of electronic spin on the dispersion curves. Clear signs of band folding are found in the c-direction electronic dispersions both below and above $E_F$. All these features harmonize with prior research [19,30].

## b) Density of States

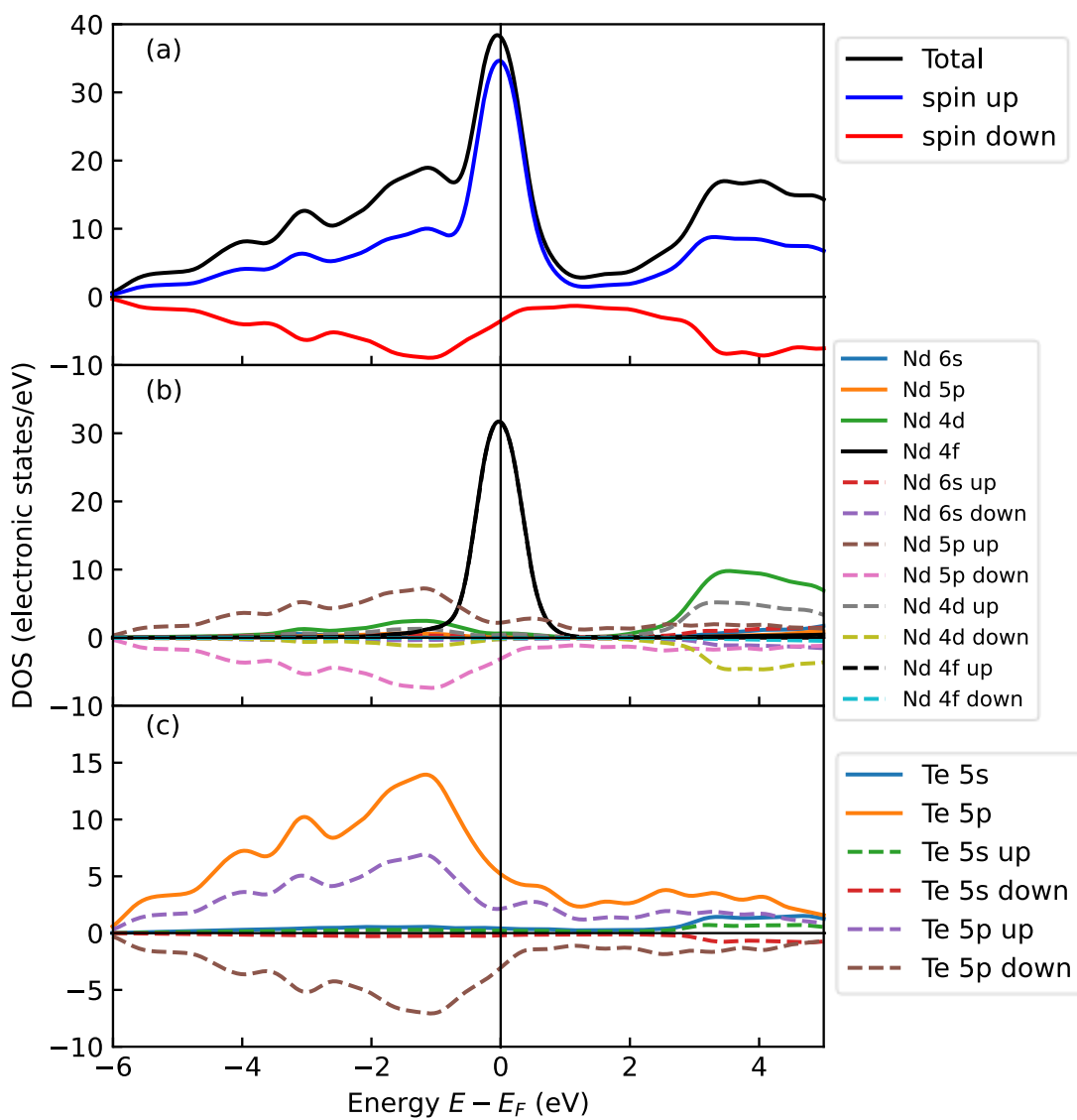

Figure 4: (a) Total density of states for NdTe$_3$, and partial density of states for (b) Nd and (c) Te atoms.

The total density of states (TDOS) along with the atom-resolved partial density of states (PDOS) as functions of energy E-$E_F$, are shown in Fig. 4. $E_F$ is represented by the vertical line at 0 eV. It is seen from Fig. 4(b), that contribution from Nd 4$f$ electrons is the largest in $E_F$, which is due to the less dispersive bands crossing $E_F$ in the BS. Apart from that, Te 5$p$ electrons have significant contribution in $E_F$, which are the dispersive bands crossing $E_F$ [Fig. 4(c)]. Various authors predicted that, contrary to our results, Te 5$p$ orbitals would dominate $E_F$ instead of Nd 4$f$, since electric conductivity is driven by Te planes. But previous theoretical calculation for NdTe$_3$ agrees well with ours. This might be because of Nd 4$f$ electrons being localized and hybridization between Nd 4$f$ and Te 5$p$ orbitals [31]. Besides the peak at $E_F$, another broad peak close to 4 eV is apparent in TDOS. The gap between these two peaks might be ascribed to the suppression in the TDOS due to the presence of an unconventional CDW. The peak in the TDOS at $E_F$ and its large value are indicative of electronic/magnetic instability. Such electronic systems show diverse electronic ground states due to small perturbations.

c) Fermi Surface

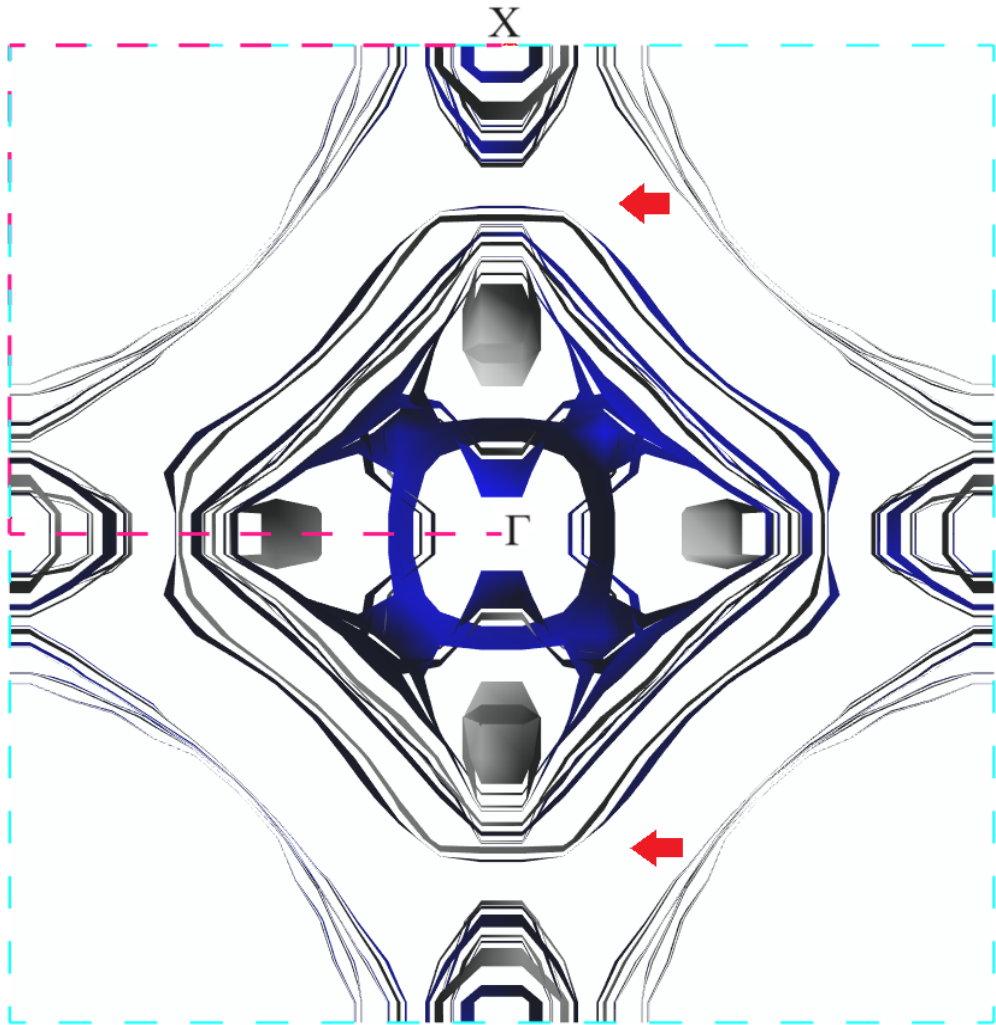

Figure 5: Fermi Surfaces of NdTe$_3$ for the 19 bands that cross the Fermi level. The dashed lines enclose the Brillouin zone.

Features unique to CDW materials are captured by their Fermi surfaces (FS). The FS plot inside the Brillouin zone (BZ) of NdTe$_3$ is shown in Fig. 5, for all the 19 bands that cross $E_F$. A diamond shaped pocket around Γ-point can be seen in the figure, as well as four "outer" pockets, only parts of which reside in the first BZ. These pockets are observed in various ARPES and QO studies of $R$Te$_3$ compounds including NdTe$_3$, as mentioned in the introduction of this paper. Quasi one-dimensional $p_x$ and $p_y$ orbitals in each square Te layer create the pockets inside their two-dimensional (2D) BZ [12,30]. All of these surfaces which create the pockets look split, which is called "bilayer splitting". This occurs because of the interplay between two Te layers within the structural unit [11,17,77]. Since the unit cell of three-dimensional (3D) crystal structure is larger than that of 2D square Te structure, the corresponding smaller 3D BZ is projected onto the 2D BZ, this makes FS elements "fold" into this smaller BZ. Shadow bands are signatures of strong electronic correlations which lead to shadow FS elements. This is intimately connected to the nesting of FS. Nesting wave vectors give the periodicity in the CDW. A CDW gap can be seen in the region between the two red arrows (Fig. 5). The gap originates from the interaction between the original bands and shadow bands. Change in FS of NdTe$_3$ induced by temperature and field is indicated by Dalgaard et al. [16]. We refer to Brouet et al. and Chikina et al. for a more comprehensive discussion on the nesting driven CDW [12,30].

### F. Optical Properties

In this section, various frequency/energy dependent optical functions have been calculated and presented. Fig. 6(a-c) depicts both the real and imaginary parts of the dielectric constants $\varepsilon_1(\omega)$ and $\varepsilon_2(\omega)$, respectively, the real part of the refractive index $n(\omega)$, the extinction coefficient $k(\omega)$, the real and imaginary parts of the optical conductivity $\sigma_1(\omega)$ and $\sigma_2(\omega)$, respectively, while fig.6 (d-f) presents reflectivity $R(\omega)$, the absorption coefficient $\alpha(\omega)$, and the loss function $L(\omega)$, respectively. Incident photon energy range was set to be 20 eV with electric field polarization vectors along [100] and [001] directions. Since the in-plane lattice parameters are very close, the corresponding anisotropy in optical functions is barely noticeable, and therefore, one of the directions among them is omitted. A Drude damping of 0.05 eV and a Gaussian smearing of 0.5 eV have been used while calculating the aforementioned optical functions.

In Fig. 6(a), it is seen that both real (Re) and imaginary (Im) parts of the complex dielectric function fall to zero approximately at 25 eV. Therefore, the plasma frequency of NdTe$_3$ should be 25eV, that is, for photons above this frequency the material is transparent to electromagnetic waves. The real part of dielectric function indicates polarizability; the imaginary part signifies loss. The real part $n(\omega)$, of the refractive index signifies the phase velocity of electromagnetic wave, while the imaginary part $k(\omega)$, alternatively extinction coefficient, is related to attenuation of wave.

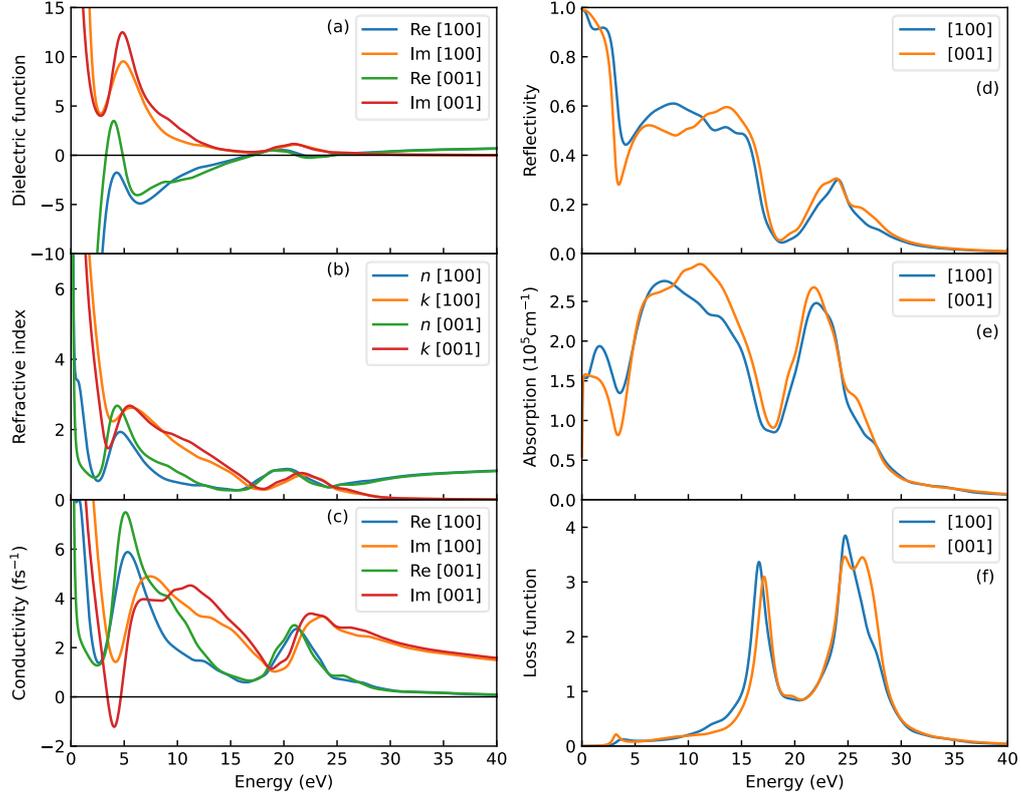

Figure 6: (a) Dielectric function (real and imaginary), (b) refractive index (real and imaginary), (c) optical conductivity (real and imaginary), (d) reflectivity, (e) absorption coefficient, and (f) loss function as functions of energy for two different polarization directions.

Metallic nature of NdTe$_3$ was discussed in the electronic band structure section. Fig. 6(c) reaffirms this, where finite value of conductivity at zero photon energy is noticed. So does the reflectivity of unity in Fig. 6(d) and non-zero value of absorption in Fig. 6(e) at zero photon energy. This metallic behavior in conductivity and reflectivity was reported experimentally as well [36]. Across the visible region, the reflectivity is very large for NdTe$_3$ in Fig. 6(d), which means that this material might be used in devices where high reflectivity is needed. Absorption of NdTe$_3$ in Fig. 6(e) is high in the energy range 6-15 eV, indicating its possible usage as an ultraviolet light absorber. The absorption sharply reaches to very close to zero around plasma frequency 25 eV. In Fig. 6(f), the peaks of the loss function, which are the plasma frequencies, occur at 17 eV and 25eV. The multiple plasma peaks imply that there are multiple energies for plasma oscillations. This is probably a multiband effect where charge carriers have different effective masses.

We can also notice two peaks in the real part of the optical conductivity spectrum at lower frequency range; one at 0 eV and another at around 4 eV, just like the density of states shown in Fig. 4(a). The former peak is called the Drude peak and the latter is a single particle excitation peak. The Drude peak indicates metallic conduction due to free charge carriers [36]. The second peak is due to the charge excitation across the CDW gap into a single particle state, which is a signature of the CDW phase [78]. The energy difference between the two peaks is associated with the CDW gap. It is also interesting to note that, despite having highly layered structural, elastic, and electronic features, the optical parameters' spectra are comparatively isotropic with respect to the polarization direction of the electric field associated with the incident light.

## IV.  Conclusions

We have explored the structural, elastic, thermomechanical and optoelectronic properties of $NdTe_3$ exploiting Density functional theory. Our optimized structural features of this layered compound are congruous with earlier works. According to the calculated elastic constants, this material is found to be mechanically stable and its ability to resist mechanical stress in the in-plane directions is stronger than that of the out-of-plane direction. The isotropic shear and bulk modulus and hardness indices suggest its soft and damage tolerant nature, while Pugh's and Poisson's ratio convey that it is a ductile material. The machinability of $NdTe_3$ is very good as well. Comprehensive investigation points out its high elastic anisotropy. When it comes to thermomechanical features, the calculated Debye temperature once again proves softness of the material, and along with low melting temperature, it implies weak interatomic bond strength. We also report low minimum thermal conductivity, which makes it a prospective thermal barrier coating material (if the operating temperature is not very high). To the best of our knowledge, none of these parameters were experimentally examined. The band structure of $NdTe_3$ reveals highly directional metallic character, with effective mass of carriers being very low in the in-plane directions. Signs of band folding in the out-of-plane direction are found as well. In the density of states, a charge density wave gap of 4 eV can be noticed. The Fermi surface shows a diamond shaped pocket around Γ-point along with four other pockets in the Brillouin zone, consistent with earlier works. They also exhibit nesting, bilayer splitting and other related phenomena. The optical conductivity, reflectivity and absorption provide confirmation that the material is metallic. The optical spectra reveal notable reflectivity in the visible region, suggesting potential application as a solar reflector. Simultaneously, high absorption in the ultraviolet region indicates the material's potential efficacy as an ultraviolet radiation absorber. We can notice multiple energies for plasma oscillations and proof of charge density wave gap in the optical functions.

**Data availability**

The data sets generated and/or analyzed in this study are available from the corresponding author on reasonable request.

**Author contributions**

T. C. performed the theoretical calculations, contributed to the analysis and draft manuscript writing. B. R. R. performed the theoretical calculations, contributed to the analysis, and contributed to manuscript writing. I. M. S. contributed to the analysis and manuscript writing. S. H. N. supervised the project, analyzed the results and finalized the manuscript. All the authors reviewed the manuscript.

**Competing Interests**

The authors declare no competing interests.

**References**

[1] S. Sarkar et al., *Charge Density Wave Induced Nodal Lines in LaTe₃*, Nat Commun **14**, 3628 (2023).
[2] Y. Chen et al., *Raman Spectra and Dimensional Effect on the Charge Density Wave Transition in GdTe₃*, Appl Phys Lett **115**, 151905 (2019).
[3] A. Fang, J. A. W. Straquadine, I. R. Fisher, S. A. Kivelson, and A. Kapitulnik, *Disorder-Induced Suppression of Charge Density Wave Order: STM Study of Pd-Intercalated ErTe₃*, Phys Rev B **100**, 235446 (2019).
[4] A. Zong et al., *Evidence for Topological Defects in a Photoinduced Phase Transition*, Nat Phys **15**, 27 (2019).
[5] E. Lee et al., *The 7 × 1 Fermi Surface Reconstruction in a Two-Dimensional f-Electron Charge Density Wave System: PrTe₃*, Sci Rep **6**, 30318 (2016).
[6] S. Lei et al., *High Mobility in a van Der Waals Layered Antiferromagnetic Metal*, Sci Adv **6**, 6407 (2020).
[7] A. Pariari, S. Koley, S. Roy, R. Singha, M. S. Laad, A. Taraphder, and P. Mandal, *Interplay between Charge Density Wave Order and Magnetic Field in the Nonmagnetic Rare-Earth Tritelluride LaTe₃*, Phys Rev B **104**, 155147 (2021).
[8] D. A. Zocco, J. J. Hamlin, K. Grube, J. H. Chu, H. H. Kuo, I. R. Fisher, and M. B. Maple, *Pressure Dependence of the Charge-Density-Wave and Superconducting States in GdTe₃, TbTe₃, and DyTe₃*, Phys Rev B **91**, 205114 (2015).
[9] J. J. Hamlin, D. A. Zocco, T. A. Sayles, M. B. Maple, J. H. Chu, and I. R. Fisher, *Pressure-Induced Superconducting Phase in the Charge-Density-Wave Compound Terbium Tritelluride*, Phys Rev Lett **102**, 177002 (2009).
[10] K. Yumigeta, Y. Qin, H. Li, M. Blei, Y. Attarde, C. Kopas, and S. Tongay, *Advances in Rare-Earth Tritelluride Quantum Materials: Structure, Properties, and Synthesis*, Advanced Science **8**, 2004762 (2021).
[11] G. H. Gweon, J. D. Denlinger, J. A. Clack, J. W. Allen, C. G. Olson, E. Di Masi, M. C. Aronson, B. Foran, and S. Lee, *Direct Observation of Complete Fermi Surface, Imperfect Nesting, and Gap Anisotropy in the High-Temperature Incommensurate Charge-Density-Wave Compound SmTe₃*, Phys Rev Lett **81**, 886 (1998).
[12] V. Brouet et al., *Angle-Resolved Photoemission Study of the Evolution of Band Structure and Charge Density Wave Properties in RTe₃ (R=Y, La, Ce, Sm, Gd, Tb, and Dy)*, Phys Rev B **77**, 235104 (2008).
[13] H. Komoda, T. Sato, S. Souma, T. Takahashi, Y. Ito, and K. Suzuki, *High-Resolution Angle-Resolved Photoemission Study of Incommensurate Charge-Density-Wave Compound CeTe₃*, Phys Rev B Condens Matter Mater Phys **70**, 195101 (2004).
[14] N. Ru, R. A. Borzi, A. Rost, A. P. Mackenzie, J. Laverock, S. B. Dugdale, and I. R. Fisher, *De Haas-van Alphen Oscillations in the Charge Density Wave Compound Lanthanum Tritelluride LaTe₃*, Phys Rev B **78**, 045123 (2008).
[15] M. Watanabe et al., *Shubnikov-de-Haas Oscillation and Possible Modification of Effective Mass in CeTe₃ thin Films*, AIP Adv **11**, 015005 (2021).
[16] K. J. Dalgaard, S. Lei, S. Wiedmann, M. Bremholm, and L. M. Schoop, *Anomalous Shubnikov-de Haas Quantum Oscillations in Rare-Earth Tritelluride NdTe₃*, Phys Rev B **102**, (2020).

[17] J. Laverock, S. B. Dugdale, Z. Major, M. A. Alam, N. Ru, I. R. Fisher, G. Santi, and E. Bruno, *Fermi Surface Nesting and Charge-Density Wave Formation in Rare-Earth Tritellurides*, Phys Rev B **71**, 085114 (2005).

[18] A. Kogar et al., *Light-Induced Charge Density Wave in LaTe$_3$*, Nature Physics 2019 16:2 **16**, 159 (2020).

[19] P. Walmsley, S. Aeschlimann, J. A. W. Straquadine, P. Giraldo-Gallo, S. C. Riggs, M. K. Chan, R. D. McDonald, and I. R. Fisher, *Magnetic Breakdown and Charge Density Wave Formation: A Quantum Oscillation Study of the Rare-Earth Tritellurides*, Phys Rev B **102**, 045150 (2020).

[20] Y. Wang et al., *Axial Higgs Mode Detected by Quantum Pathway Interference in RTe$_3$*, Nature **606**, 896 (2022).

[21] B. K. Norling and H. Steinfink, *The Crystal Structure of Neodymimum Tritelluride*, Inorg Chem **5**, 1488 (1996).

[22] C. Malliakas, S. J. L. Billinge, J. K. Hyun, and M. G. Kanatzidis, *Square Nets of Tellurium: Rare-Earth Dependent Variation in the Charge-Density Wave of RETe$_3$ (RE = Rare-Earth Element)*, J Am Chem Soc **127**, 6510 (2005).

[23] Y. Iyeiri, T. Okumura, C. Michioka, and K. Suzuki, *Magnetic Properties of Rare-Earth Metal Tritellurides RTe$_3$ (R=Ce,Pr,Nd,Gd,Dy)*, Phys Rev B **67**, 1444171 (2003).

[24] N. Ru, J.-H. Chu, and I. R. Fisher, *Magnetic Properties of the Charge Density Wave Compounds RTe$_3$, R = Y, La, Ce, Pr, Nd, Sm, Gd, Tb, Dy, Ho, Er & Tm*, Phys Rev B **78**, 012410 (2008).

[25] L. M. Schoop, F. Pielnhofer, and B. V. Lotsch, *Chemical Principles of Topological Semimetals*, Chemistry of Materials **30**, 3155 (2018).

[26] E. DiMasi, M. C. Aronson, J. F. Mansfield, B. Foran, and S. Lee, *Chemical Pressure and Charge-Density Waves in Rare-Earth Tritellurides*, Phys Rev B **52**, 14516 (1995).

[27] N. Ru, C. L. Condron, G. Y. Margulis, K. Y. Shin, J. Laverock, S. B. Dugdale, M. F. Toney, and I. R. Fisher, *Effect of Chemical Pressure on the Charge Density Wave Transition in Rare-Earth Tritellurides RTe$_3$*, Phys Rev B **77**, 035114 (2008).

[28] A. Sacchetti et al., *Pressure-Induced Quenching of the Charge-Density-Wave State in Rare-Earth Tritellurides Observed by x-Ray Diffraction*, Phys Rev B **79**, 201101(R) (2009).

[29] A. Banerjee, Y. Feng, D. M. Silevitch, J. Wang, J. C. Lang, H. H. Kuo, I. R. Fisher, and T. F. Rosenbaum, *Charge Transfer and Multiple Density Waves in the Rare Earth Tellurides*, Phys Rev B **87**, 155131 (2013).

[30] A. Chikina, H. Lund, M. Bianchi, D. Curcio, K. J. Dalgaard, M. Bremholm, S. Lei, R. Singha, L. M. Schoop, and P. Hofmann, *Charge Density Wave-Generated Fermi Surfaces in NdTe$_3$*, Phys Rev B **107**, L161103 (2023).

[31] Y. Hong, Q. Wei, X. Liang, and W. Lu, *Origin and Strain Tuning of Charge Density Wave in LaTe$_3$*, Physica B Condens Matter **639**, 413988 (2022).

[32] M. D. Johannes and I. I. Mazin, *Fermi Surface Nesting and the Origin of Charge Density Waves in Metals*, Phys Rev B **77**, 165135 (2008).

[33] C. Haas, *Spin-Disorder Scattering and Magnetoresistance of Magnetic Semiconductors*, Physical Review **168**, 531 (1968).

[34] Ł. Gondek1, A. Szytuła, D. Kaczorowski, A. Szewczyk, M. Gutowska and P. Piekarz, *Multiple magnetic phase transitions in Tb$_3$Cu$_4$Si$_4$*, J Phys: Condens Matter **19**, 246225 (2007).

[35] M. Rotter, M. Tegel, D. Johrendt, I. Schellenberg, W. Hermes, and R. Pöttgen, *Spin-Density-Wave Anomaly at 140 K in the Ternary Iron Arsenide BaFe$_2$As$_2$*, Phys Rev B **78**, 020503 (2008).

[36] A. Sacchetti, L. Degiorgi, T. Giamarchi, N. Ru, and I. R. Fisher, *Chemical Pressure and Hidden One-Dimensional Behavior in Rare-Earth Tri-Telluride Charge-Density Wave Compounds*, Phys Rev B **74**, 125115 (2006).

[37] M. Lavagnini et al., *Evidence for Coupling between Charge Density Waves and Phonons in Two-Dimensional Rare-Earth Tritellurides*, Phys Rev B **78**, 201101(R) (2008).

[38] A. Sacchetti, E. Arcangeletti, A. Perucchi, L. Baldassarre, P. Postorino, S. Lupi, N. Ru, I. R. Fisher, and L. Degiorgi, *Pressure Dependence of the Charge-Density-Wave Gap in Rare-Earth Tritellurides*, Phys Rev Lett **98**, 026401 (2007).

[39] J. B. He et al., *Superconductivity in Pd-Intercalated Charge-Density-Wave Rare Earth Poly-Tellurides RETe$_n$*, Supercond Sci Technol **29**, 065018 (2016).

[40] J. Gjerde and R. A. Jishi, *Hyperbolic Behavior and Antiferromagnetic Order in Rare-Earth Tellurides*, Crystals **12**, 1839 (2022).

[41] J. Kopaczek et al., *Experimental and Theoretical Studies of the Surface Oxidation Process of Rare-Earth Tritellurides*, Adv Electron Mater **9**, 2201129 (2023).

[42] W. Kohn and L. J. Sham, *Self-Consistent Equations Including Exchange and Correlation Effects*, Phys Rev **140**, A1133 (1965).

[43] J. P. Perdew and A. Zunger, *Self-Interaction Correction to Density-Functional Approximations for Many-Electron Systems*, Phys Rev B **23**, 5048 (1981).

[44] S. J. Clark, M. D. Segall, C. J. Pickard, P. J. Hasnip, M. I. J. Probert, K. Refson, and M. C. Payne, *First Principles Methods Using CASTEP*, Zeitschrift Fur Kristallographie **220**, 567 (2005).

[45] D. R. Hamann, M. Schlüter, and C. Chiang, *Norm-Conserving Pseudopotentials*, Phys Rev Lett **43**, 1494 (1979).

[46] D. D. Koelling and B. N. Harmon, *A Technique for Relativistic Spin-Polarised Calculations*, J Phys C: Solid State Phys **10**, 3107 (1977).

[47] T. H. Fischer and J. Almlöf, *General Methods for Geometry and Wave Function Optimization*, J Phys Chem **96**, 9768 (1992).

[48] H. J. Monkhorst and J. D. Pack, *Special Points for Brillouin-Zone Integrations*, Phys Rev B **13**, 5188 (1976)

[49] Z. Xu et al., *Molecular Beam Epitaxy Growth and Electronic Structures of Monolayer GdTe$_3$*, Chinese Physics Letters **38**, 077102 (2021).

[50] M.I. Naher and S.H Naqib, *An ab-initio study on structural, elastic, electronic, bonding, thermal, and optical properties of topological Weyl semimetal TaX (X = P, As)*, Sci Rep **11**, 5592 (2021)

[51] M.A. Hadi, N. Kelaidis, S.H. Naqib, A. Chroneos, A.K.M.A. Islam, *Mechanical behaviors, lattice thermal conductivity and vibrational properties of a new MAX phase Lu$_2$SnC*, J Phys Chem Solids **129**, 162 (2019)

[52] B. R. Rano, I. M. Syed, S.H. Naqib, *Ab initio approach to the elastic, electronic, and optical properties of MoTe$_2$ topological Weyl semimetal,* J Alloys Compd, **829**, 154522 (2020)

[53] N. S. Khan, B. R. Rano, I. M. Syed, R.S. Islam, S.H. Naqib, *First-principles prediction of pressure dependent mechanical, electronic, optical, and superconducting state properties of NaC$_6$: A potential high-Tc superconductor*, Results Phys **33**, 105182 (2022)

[54] S. Barua, B. R. Rano, I. M. Syed, S.H. Naqib, *An ab initio approach to understand the structural, thermophysical, electronic, and optical properties of binary silicide SrSi$_2$: A double Weyl semimetal*, Result Phys **42**, 105973 (2022)

[55] S. Azad, B. R. Rano, I. M. Syed, S.H. Naqib, *A comparative study of the physical properties of layered transition metal nitride halides MNCl (M = Zr, Hf): DFT based insights,* Phys Scr **98**, 115982 (2023)

[56] M. Born, *On the Stability of Crystal Lattices. I*, Mathematical Proceedings of the Cambridge Philosophical Society **36**, 160 (1940).

[57] F. Mouhat and F. X. Coudert, *Necessary and Sufficient Elastic Stability Conditions in Various Crystal Systems*, Phys Rev B **90**, 224104 (2014).

[58] P. Ravindran, L. Fast, P. A. Korzhavyi, B. Johansson, J. Wills, and O. Eriksson, *Density Functional Theory for Calculation of Elastic Properties of Orthorhombic Crystals: Application to TiSi$_2$*, J Appl Phys **84**, 4891 (1998).

[59] W. Voigt, *Lehrbuch Der Kristallphysik*, Teubner Verlag, Leipzig, 1928.

[60] A. Reuss, *Berechnung Der Fließgrenze von Mischkristallen Auf Grund Der Plastizitätsbedingung Für Einkristalle .*, ZAMM - Journal of Applied Mathematics and Mechanics / Zeitschrift Für Angewandte Mathematik Und Mechanik **9**, 49 (1929).

[61] R. Hill, *The Elastic Behaviour of a Crystalline Aggregate*, Proceedings of the Physical Society. Section A **65**, 349 (1952).

[62] G. N. Greaves, A. L. Greer, R. S. Lakes, and T. Rouxel, *Poisson's Ratio and Modern Materials*, Nat Mater **10**, 823 (2011).

[63] O. L. Anderson and H. H. Demarest, *Elastic Constants of the Central Force Model for Cubic Structures: Polycrystalline Aggregates and Instabilities*, J Geophys Res **76**, 1349 (1971).

[64] Z. Sun, D. Music, R. Ahuja, and J. M. Schneider, *Theoretical Investigation of the Bonding and Elastic Properties of Nanolayered Ternary Nitrides*, Phys Rev B **71**, 193402 (2005).

[65] A. L. Ivanovskii, *Hardness of Hexagonal AlB$_2$-like Diborides of s, p and d Metals from Semi-Empirical Estimations*, Int J Refract Metals Hard Mater **36**, 179 (2013).

[66] X. Q. Chen, H. Niu, D. Li, and Y. Li, *Modeling Hardness of Polycrystalline Materials and Bulk Metallic Glasses*, Intermetallics (Barking) **19**, 1275 (2011).

[67] X. Jiang, J. Zhao, and X. Jiang, *Correlation between Hardness and Elastic Moduli of the Covalent Crystals*, Comput Mater Sci **50**, 2287 (2011).

[68] M. I. Naher, M. A. Ali, M. M. Hossain, M. M. Uddin, and S. H. Naqib, *A Comprehensive Ab-Initio Insights into the Pressure Dependent Mechanical, Phonon, Bonding, Electronic, Optical, and Thermal Properties of CsV$_3$Sb$_5$ Kagome Compound*, Results Phys **51**, 106742 (2023).

[69] C. M. Kube and M. De Jong, *Elastic Constants of Polycrystals with Generally Anisotropic Crystals*, J Appl Phys **120**, (2016).

[70] S. I. Ranganathan and M. Ostoja-Starzewski, *Universal Elastic Anisotropy Index*, Phys Rev Lett **101**, 055504 (2008).

[71] V. Milman and M. C. Warren, *Elasticity of Hexagonal BeO*, J of Phys: Condens Matter **13**, 241 (2001).

[72] R. Gaillac, P. Pullumbi, and F. X. Coudert, *ELATE: An Open-Source Online Application for Analysis and Visualization of Elastic Tensors*, J Phys: Condens Matter **28**, 275201 (2016).

[73] M. E. Fine, L. D. Brown, and H. L. Marcus, *Elastic Constants versus Melting Temperature in Metals*, Scripta Metallurgica **18**, 951 (1984).

[74]     G. A. Slack, *The Thermal Conductivity of Nonmetallic Crystals*, J Phys C: Solid State Phys **34**, 1 (1979).

[75]     D. R. Clarke, *Materials Selection Guidelines for Low Thermal Conductivity Thermal Barrier Coatings*, Surf Coat Technol **163–164**, 67 (2003).

[76]     E. DiMasi, B. Foran, M. C. Aronson, and S. Lee, *Quasi-Two-Dimensional Metallic Character of $Sm_2Te_5$ and $SmTe_3$*, Chem Mater **6**, 1867 (1994).

[77]     V. Brouet, W. L. Yang, X. J. Zhou, Z. Hussain, N. Ru, K. Y. Shin, I. R. Fisher, and Z. X. Shen, *Fermi Surface Reconstruction in the CDW State of $CeTe_3$ Observed by Photoemission*, Phys Rev Lett **93**, 126405 (2004).

[78]     P. A. Lee, T. M. Rice and P. W. Anderson, *Conductivity from Charge or Spin Density Waves*, Solid State Commun **14**, 703 (1974).